# An Iterative Heuristic Method to Determine Radial Topology for Distribution System Restoration

Jiayu Liu, Qiqi Zhang
State Grid Shanghai Electrical
Power Company Research Institute
Shanghai, China

Jiaxu Li, Ying Wang
Beijing Jiaotong University
School of Electrical Engineering
Beijing, China
lijiaxu@bjtu.edu.cn

*Abstract*—Coordinating multiple local power sources can restore critical loads after the major outages caused by extreme events. A radial topology is needed for distribution system restoration, while determining a good topology in real-time for online use is a challenge. In this paper, a graph theory-based heuristic considering power flow state is proposed to fast determine the radial topology. The loops of distribution network are eliminated by iteration. The proposed method is validated by one snapshot and multi-period critical load restoration models on different cases. The case studies indicate that the proposed method can determine radial topology in a few seconds and ensure the restoration capacity.

## I. INTRODUCTION

With the increasing frequency of power outages caused by extreme events, the power system resilience has become a hotspot in power industry and academic [1]. The distribution system is usually disconnected from the upstream transmission system after outages. Using multiple local sources for service restoration to critical loads is an effective method to reduce the loss of outages and enhance resilience [2].

The distribution system is designed as a meshed network but operates radially [3]. A good radial topology is needed to maximize the restoration capability of the system. The restoration capability can be affected by the topology as power flow distribution will be different for systems with different topologies.

Several scholars seek mathematical programming methods to solve the distribution system restoration (DSR) problem. Maintaining radial topology after restoration is considered as a constraint in much former work. The spanning tree (ST) constraint proposed in [4] has been widely used. Reference [5] formulate the DSR problem as a mixed-integer second-order conic program (MISOCP), in which the radial topology constraints are represented by ST constraint. A mixed-integer linear program (MILP) is proposed to determine the restoration status of loads and lines [6]. However, reference [3] found that the ST constraint could not guarantee the obtained topology to be radial in some scenarios based on a large number of tests. Reference [7] points that the ST constraint may result in a non-connected graph with a loop, but it does not analyze the detailed reason. In addition, due to the nonconvex power flow constraint and the integer variables including loads and lines status, the DSR model is hard to solve. Even though the nonconvex power flow constraint can be formulated and relaxed as convex constraints, the 0-1 binary variables are not easy to handle. The computation burden will be heavy when the number of integer variables is huge, which cannot satisfy the online requirement.

The heuristic methods are also widely used in the DSR problem. Based on the restoration criteria and operation rules, the restoration result is determined [8]. However, the solution obtained by heuristic algorithm may be very far from the global optimal solution [9]. In authors' previous work, the DSR problem is solved by combining the heuristic and mathematic programming methods [10]. The restoration result is obtained in two stages. In stage one, the radial topology (line status variable) is determined by the heuristic method. In stage two, the load status variables are handled. However, the heuristic method for radial topology determination in [10] may reduce the restoration capacity.

In this paper, a heuristic method to fast determine the radial topology for the DSR problem is proposed. The main contributions include:

1) A topology determining heuristic method based on graph theory is proposed, in which the power flow state is considered.

2) The benefit of using the heuristic method to determine radial topology in both one-snap shot and multi-period DSR problems are analyzed, validated by modified IEEE 32-node system, IEEE 123-node system, and 62-node distribution system.

The remainder of this paper is organized as follows. Section II describes the DSR problem. Section III formulates the heuristic method. Numerical results are presented in Section IV. Section V concludes the study.

## II. PROBLEM DESCRIPTION

The restoration status of loads and lines, as well as power source outputs, are needed to be determined in the DSR problem. Due to limited generation resources, critical loads like lifeline facilities are restored with high priority. The

This work is supported by State-grid Shanghai Municipal Electric Power Company Project (No. B30940190000).

power flow constraint, operational constraint, and radial constraint after restoration are needed to be followed.

Considering the outage usually lasts for several hours, the limited generator resources should be appropriately allocated to serve critical loads for a long duration [11]. To consider the operation during the outage, the one snapshot DSR model can be extended to a multi-period DSR model. The objective is modified as the cumulative restored time of critical loads. The mentioned constraints are extended to multi-period. The resource constraint and SOC constraint are added. The details of one snapshot DSR model and multi-period DSR model are presented in our previous work [10],[11].

The purpose of this study is to determine the restoration status of lines in two DSR models. A distribution network is composed of buses and lines connecting these buses. Generally, the distribution network can be deemed a connected meshed graph $G=\langle N,E\rangle$, where $N$ is the set of all buses and $E$ is the set of all available lines. Assumed that all tie lines are connected, graph G contains several cycles. We need to find a tree $G'=\langle N,E'\rangle$ by cutting $|E|-|N|+|S|$ lines, where $S$ is the set of power sources and the number is denoted as $|S|$. Specially, we coordinate multiple local sources as a large island, which means $|S|=1$ [10].

## III. Proposed Algorithm

In this section, an iterative heuristic method to determine radial topology for the DSR problem is proposed.

### A. The Iterative Heuristic Method

The iterative heuristic (IH) method is to cut one loop line in each iteration based on the power flow state, which converges when no loop exists in the graph. In each iteration, two basic steps are involved:

Step 1: Solve a critical load restoration optimization model for meshed network G, which will be presented in detail in the next part.

Step 2: Open loop-lines to eliminate loops based on the value of the active power of lines.

The iteration times is equal to the number of meshes. A diagram of the iteration is shown in Fig. 1.

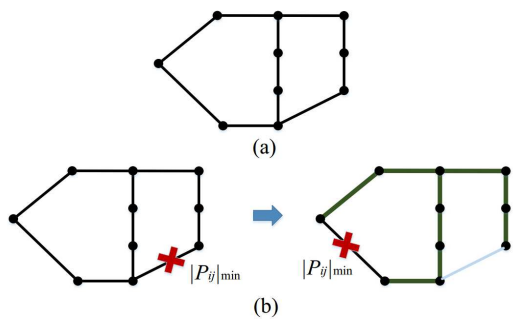

Fig. 1. A diagram of the iterative heuristic method. (a) A sample of distribution network; (b) The iterative process.

We cut the line with the minimum active power in the loop lines because the line is the least important for transmitting the active power with the restoration objective. This idea is inspired by the feeder reconfiguration method in [12] where the line that receives positive real power from both sides in the path connecting two feeders is opened. The line receiving active power from both sides is exactly the line with the minimum active power.

The pseudo-code of the iterative heuristic is as follows. It returns the radial topology $E'$ and the set of cut lines $E_c$

| **Algorithm** Iterative Heuristic |
|---|
| 1  Input the network with all lines connected $G=\langle N,E\rangle$ |
| 2  The number of loops $n_l \leftarrow |E|-|N|+1$ |
| 3  Initiate $E'\leftarrow E$, $E_c \leftarrow \varnothing$ |
| 4  **while** $n_l \geq 1$ |
| 5      Solve the restoration model for meshed network $G=\langle N,E'\rangle$ |
| 6      Find the set of loop lines $E_l$ using the depth-first search in [13] |
| 7      Find $i\to j\in E_l$ with the minimum positive active power |
| 8      $E_c \leftarrow E_c \bigcup i\to j$ |
| 9      $E'\leftarrow E'\setminus i\to j$ |
| 10     $n_l \leftarrow n_l -1$ |
| 11 **end** |
| 12 **Return** sets $E'$ and $E_c$ |

The performance of the algorithm relies on the algorithms solving the restoration model in line 5 and finding the set of loop lines in line 6. The restoration model is a quadratic program (QP) as shown in the next part. To solve QP, one can find the optimum using algorithms such as the interior algorithm in polynomial time [14]. The depth-first search can also terminate in polynomial time [13]. Therefore, the IH method can also obtain the solution in polynomial time.

### B. Critical Load Restoration Model for Meshed Network

As mentioned, the topology will affect the power flow distribution. The number of loads restored is usually mainly yielded by power capacity constraints rather than voltage constraints, as the sources are geographically dispersed and they can support the voltage. The bus voltage does not monotonically decrease along with the lines from substations to loads, so the upper or lower voltage limits are not likely to be met. Based on the analysis, two assumptions are made to model the restoration problem for fast determining the topology: 1) The voltage magnitudes for all buses are assumed to be equal, i.e., $V_{rate}=1\text{p.u.}$; 2) The load in each node can be partially restored. The model **CLR-mesh** is as follows.

**CLR-mesh**

$$\max f = N_{load} - w_0 P_{loss}$$
$$= \sum_{i \in L} w_i \gamma_i - w_0 \sum_{i \to j \in E} \frac{R_{ij}(P_{ij}^2 + Q_{ij}^2)}{(V_{rate})^2} \quad (1)$$

over $P_{ij}, Q_{ij} \in R$, for $i \to j \in E$; $p_{gen,i}, q_{gen,i} \in R$, for $i \in S$; $\gamma_i \in [0,1]$, for $i \in L$

s.t. $\sum_{k:k \to i} P_{ki} - \gamma_i p_{load,i} + p_{gen,i} = \sum_{j:i \to j} P_{ij}, \forall i \in N \quad (2)$

$\sum_{k:k \to i} Q_{ki} - \gamma_i q_{load,i} + q_{gen,i} = \sum_{j:i \to j} Q_{ij}, \forall i \in N \quad (3)$

$p_{gen,i} \leq p_{i,\max}, q_{gen,i} \leq q_{i,\max}, \forall i \in S \quad (4)$

$|P_{ij}| \leq P_{ij,\max}, \forall i \to j \in E \quad (5)$

where L, S are the sets of all load buses and source buses, respectively; $P_{ij}, Q_{ij}$ are the variables indicating active and reactive power of line $i \to j$, respectively; $p_{gen,i}$ and $q_{gen,i}$ are the active and reactive power generated by source $i$, respectively; $\gamma_i$ is the variable indicating load status; $w_i, w_0, p_{load,i}, q_{load,i}, R_{ij}, P_{ij,\max}, p_{i,\max}, q_{i,\max}$ are the constants indicating the weighting factor of loads, weighting factor balancing two objectives, active power demand, and reactive power demand of load $i$, resistance and thermal limit of line $i \to j$, and the maximum active and reactive power of source $i$, respectively.

In (1), the first term indicates the main objective, i.e., the weighted number of restored loads $N_{load}$, and the second is the power loss $P_{loss}$ of the system. The weighting factor $w_0$ is set as 0.001. Constraints (2) and (3) are the power balance constraints. Constraint (4) is the power capacity constraint for sources. Constraint (5) is the thermal limit for each line.

*C. Multi-Period Critical Load Restoration Model for Meshed Network*

Similarly, the multi-period critical load restoration model for meshed network is formulated.

**MPCLR-mesh**

$$\max f = \sum_{t \in T} \sum_{i \in L} w_i \gamma_i^t - w_0 \sum_{t \in T} \sum_{i \to j \in E} \frac{R_{ij}(P_{ij}^{t2} + Q_{ij}^{t2})}{(V_{rate})^2} \quad (6)$$

over $P_{ij}^t, Q_{ij}^t \in R$, for $i \to j \in E, t \in T$;
$p_{gen,i}^t, q_{gen,i}^t \in R$, for $i \in S, t \in T$;
$\gamma_i^t \in [0,1]$, for $i \in L, t \in T$;

s.t. $\sum_{t \in T} p_{gen,i}^t T_{int} \leq E_i, \forall i \in S / B, t \in T \quad (7)$

$SOC_i^{\min} \leq SOC_i + \rho_i \sum_{t \in T} p_{gen,i}^t T_{int} \leq SOC_i^{\max}, \forall i \in B, t \in T \quad (8)$

$\sum_{k:k \to i} P_{ki}^t - \gamma_i^t p_{load,i} + p_{gen,i}^t = \sum_{j:i \to j} P_{ij}^t, \forall i \in N, t \in T \quad (9)$

$\sum_{k:k \to i} Q_{ki}^t - \gamma_i^t q_{load,i} + q_{gen,i}^t = \sum_{j:i \to j} Q_{ij}^t, \forall i \in N, t \in T \quad (10)$

$p_{gen,i}^t \leq p_{i,\max}, q_{gen,i}^t \leq q_{i,\max}, \forall i \in S, t \in T \quad (11)$

$|P_{ij}^t| \leq P_{ij,\max}, \forall i \to j \in E, t \in T \quad (12)$

where B is the sets of buses connected energy storages. The whole time of restoration is denoted as T and is divided into several intervals. The length of each time interval is $T_{int}$. The notation with superscript $t$ indicates the corresponding variable in time $t$.

Similar to **CLR-Mesh**, the objective function of **MPCLR-Mesh** contains the primary objective and the secondary objective. The primary objective is the sum of the weighted number of restored loads in each period, and the secondary objective is the cumulative power loss in all periods. Constraints (7) and (8) are limited generator resources and SOC constraints, respectively. Constraints (9)-(12) are extended from (2)-(5).

The **CLR-Mesh** and **MPCLR-Mesh** model are convex QP, which can be solved readily by the off-the-shelf solvers.

## IV. CASE STUDY

To validate the accuracy and efficiency of the proposed method, the topologies determined by IH and mathematic programming method in CLR and MPCLR are compared.

*A. Performance of the Proposed Heuristics in CLP*

The modified 32-node [15] and IEEE 123-Node systems [10] with distributed generations (DGs) added and critical loads classified are used in this part. The numbers of switchable lines and loads for two systems are 36, 32, 124, and 85, respectively. The proposed heuristic is used to obtain radial topologies of the two systems for restoration.

To measure the quality of the proposed methods, the obtained topologies of IH, maximum spanning tree (MST)-based heuristic in [16], minimum diameter spanning tree (MDST)-based heuristic in [10], and mathematic programming method are compared. The detailed comparison procedure is as follow:

1) 300 scenarios with different locations of DGs and critical loads are generated.

2) For heuristic methods, the radial topologies after restoration of each scenario are determined using IH, MST, and MDST methods, respectively. Then, the weighted number of restored loads $N_{load}$ and power losses $P_{loss}$ are determined using the critical load restoration model **CLR-misocp** with objective function (1) in the Appendix of [10].

3) For mathematic programming methods, the MISOCP in [15] determining both line and load status is conducted, whose objective value is deemed the global optimum $f^*$. The MISOCP is solved by MOSEK and the maximum computation time is set as 100s [17].

4) The radial topology and load restoration status obtained by four methods of each scenario are compared.

An error factor $\sigma$ is defined to measure the solution quality of different methods with the global solution:

$$\sigma = \frac{|f^* - f_x|}{f^*} \quad (13)$$

where $f_x$ means the objective value obtained by topology determined by heuristic $x$. The solution is deemed a near-optimum when $\sigma \leq 10^{-4}$. The results of the solution quality are shown in Table I.

TABLE I. SOLUTION QUALITY OF THREE HEURISTICS COMPARING WITH MISOCP

| Case | Meshes | Number of scenarios | IH | MST | MDST |
|---|---|---|---|---|---|
| 32-Node | 5 | Same Topo. | 20 | 43 | 0 |
| | | Near-optimum | 300 | 300 | 234 |
| | | $N_{load} < N^*_{load}$ | 0 | 0 | 51 |
| 123-Node | 2 | Same Topo. | 20 | 19 | 1 |
| | | Near-optimum | 300 | 300 | 298 |
| | | $N_{load} < N^*_{load}$ | 0 | 0 | 2 |

In TABLE I, the entry "Same Topo." means the scenarios that heuristic $x$ obtains the same topology with MISOCP, "Near-optimum" means the scenarios that heuristic $x$ obtains near-optimal solution, and "$N_{load} < N^*_{load}$" means the scenarios that heuristic $x$ obtains less weighted number of restored loads indicating the restoration capacity is reduced by the topology.

For two systems, the restoration results of IH and MST in 300 scenarios all can obtain the near-optimal topology to ensure the restoration capability of the system. Heuristic MDST cannot guarantee the restoration capability of the system as the weighted number of loads may decrease using the topology determined. It is concluded that inappropriate topology can reduce restoration capability.

The topologies obtained by MISOCP and IH may not be the same but the restoration capability can also be retained with minor difference in power loss. Near-optimal topologies that can retain the restoration capability are acceptable in practice.

In IEEE 32-node test system, the number of the same topology scenarios MST is more than IH, while the result in IEEE 123-node system is opposite. The number of meshes has a huge impact on the restoration result. The computation time results of IH and MST are illustrated in Table II.

TABLE II. CONPUTATAION TIME OF IH, MST, AND MISOCP

| Case | Meshes | Time(s) | IH | MST | MSOCP |
|---|---|---|---|---|---|
| 32-Node | 5 | $t_{min}$ | 1.91 | 0.36 | 0.70 |
| | | $t_{max}$ | 2.57 | 0.55 | 100.00 |
| | | $t_{ave}$ | 2.02 | 0.40 | 19.90 |
| 123-Node | 2 | $t_{min}$ | 2.75 | 1.17 | 0.89 |
| | | $t_{max}$ | 3.43 | 1.70 | 100.00 |
| | | $t_{ave}$ | 2.92 | 1.43 | 74.87 |

The average computation time for IH is about 5 and 2 times longer than MST-based one for 32-Node and 123-Node system, as the numbers of iterations are 5 and 2, respectively. IH needs to 5 and 2 continuous convex optimization models, while MST only needs to solve one model. Generally, the IH and MST are both faster than MISOCP with acceptable near-optimal solutions. After topology is determined, one can use the algorithm in [10] to determine restoration status of loads within 30 seconds for the 123-Node system.

To further analyze the applicability of IH and MST, cases with large loads, low thermal limits, and high impedance indicating different types of distribution systems are tested. The results of IH and MST are similar except for the conditions with low thermal limits. The results are illustrated in Table III.

TABLE III. RESULTS OF IH AND MST FOR DISTRIBUTION SYSTEM WITH LOW THERMAL LIMITS IN 100 SCENARIOS

| Number of scenarios | IH | MST |
|---|---|---|
| Near-optimum | 100 | 64 |
| $N_{load} < N^*_{load}$ | 0 | 31 |

According to Table II and Table III, even the computation time of IH is a bit longer than MST, the IH can suit more operation conditions of distribution system.

B. Performance of the Proposed Heuristics in MPCLP

In this part, a modified 62-node distribution system [18] is used to test the optimality and computation speed of IH. 1000 scenarios with different locations of critical loads are generated.

Similar to the last part, the restoration results of MPCLR solved by four methods are compared. The MPCLR-misocp model in [18] is solved by MOSEK and the maximum computation time is set as 30min. If MPCLR-misocp problem fails to be solved after 30min, it is considered as non-convergence. During the test, we find that the number of non-convergence scenarios is 754, which means the result of mathematic programming methods may not be the global optimum. Thus, we define the maximum value of four objectives solved by three methods as the reference result $f^*_{MPCLR}$. The solution is deemed a near-optimum when $\sigma \leq 10^{-3}$. The comparison result is shown in Table IV.

TABLE IV. SOLUTION QUALITY OF IH, MDST, AND MPCLR-MISOCP

| Method | Percentage | | | Value of $\sigma$ $(10^{-3})$ | |
|---|---|---|---|---|---|
| | $\sigma = 0$ | $0 < \sigma \leq 10^{-3}$ | $\sigma > 10^{-3}$ | $\sigma_{ave}$ | $\sigma_{max}$ |
| IH | 77.5% | 18.8% | 3.7% | 2.19 | 2.47 |
| MDST | 57.7% | 31.5% | 10.8% | 5.25 | 200 |
| MPCLR-misocp | 57.7% | 17.6% | 24.7% | 3.90 | 660 |

According to Table IV, the scenarios where the results obtained by IH is the same as the reference result account for more than 75%. Although the near-optimum solution cannot be obtained by IH in a few scenarios, the distance between the result and the reference result is not large ($\sigma_{max}$). In the worst scenario, the restoration time of a 1st-level and a 2nd-level loads decreases by one period.

The number of scenarios where the radial topology obtained by MDST is the same as the reference result is 577. And the $\sigma_{max}$ is much larger than proposed method. In the worst scenario, the restoration time of four 1st-level loads decreases by two periods, as well as one 2nd-level load and one normal load decrease by one period.

The radial topology obtained by MPCLR-misocp is the worst in four methods. In the worst scenario, the restoration time of four 1st-level loads decreases by seven periods and one 2nd-level load decreases by one period. It is indicated that directly solving the MPCLR model by the off-the-shelf solver cannot ensure online decision. In addition, the computation time of the proposed method and MPCLR-misocp is shown in Table V.

TABLE V. CONPUTATION TIME OF IH AND MPCLR-MISOCP

| Case | Meshes | Time(s) | IH | MPCLR-misocp |
|---|---|---|---|---|
| 62-Node | 4 | $t_{min}$ | 16.58 | 120.47 |
| | | $t_{max}$ | 22.79 | 1633.37 |
| | | $t_{ave}$ | 18.32 | 1800 |

According to Table V, compared with the average computation time of the MPCLR-misocp, which is about 27.22 minutes, the IH methods can determine the radial topology in 25 seconds.

For the scenarios where the MPCLR-misocp model can be solved directly by MOSEK in 30 minutes, we consider that the global optimal solution is obtained. This result is considered as the global optimum. If $\sigma \leq 10^{-3}$, it is defined that the restoration result is the near-optimal solution. The results are shown in Table VI.

TABLE VI. SOLUTION QUALITY OF IH UNDER SELECTED 246 SCENARIOS

| Number of scenarios | IH |
|---|---|
| Global optimum | 177 |
| Near-optimum | 44 |
| $N_{load} < N_{load}^*$ | 25 |

From Table VI, the IH method can obtain the global optimal solution or near-optimal solution in 89.84% of scenarios. Combining solution quality presented in Table IV, the IH is obviously better than MPCLR-misocp, and it is more suitable for online application.

V. CONCLUSION

This paper proposes an iterative heuristic considering the power flow state to fast determine radial topology for service restoration. The producer is to cut one loop line in each iteration and converges until no loop exists. The case studies show that the IH can obtain near-optimal radial topology of one snapshot and multi-period critical load restoration models in seconds. Combining solution quality and speed, the IH is expected to be applied online.

*Acknowledgment*

The authors would like to thank Professor Stephen Boyd for his useful suggestions on this work.